\documentclass[preprint,pre,floatfix]{revtex4}
\usepackage[latin1]{inputenc}
\usepackage[T1]{fontenc}
\usepackage[english]{babel}
\usepackage[dvips]{graphicx}
\usepackage{amsfonts}
\usepackage{amsmath}

\newcommand{\diff}{\text{d}}

\renewcommand{\vec}[1]{\text{\boldmath$#1$}}

\newcommand{\reals}{\ensuremath{\mathbb R}}

\def\braket#1{\mathinner{\langle{#1}\rangle}}

\begin{document}

\title[On the dynamics of point vortices in an annulus]{On the dynamics of point vortices in a quantum gas confined in an annular region}

\date{\today}

\author{Markus Lakaniemi}
\affiliation{Department of Physics, 20014 University of Turku, Finland}
\email{markus.lakaniemi@utu.fi}

\begin{abstract}
The dynamics of one and two pointlike vortices in a planar quantum gas of spin-$0$ particles confined in an annular region is considered. New analytical and numerical solutions are found. The concept of stationarity radius, related to the doubly connected nature of the annulus, is defined. It is seen that the existence of these radii has great impact on the behaviour of the vortices. It is shown that, because of the existence of the stationarity radii, vortices exhibit similar behaviour regardless of the sign of their winding number. The energetically stable vortex solutions are studied qualitatively. 
\end{abstract}

\maketitle

\section{Introduction}

Several systems in low temperature physics exhibit different kinds of vortex solutions. Because these solutions have many interesting properties, vortices have been extensively studied both theoretically and experimentally \cite{Onsager(Feynman-Onsager),Feynman(Feynman-Onsager),first experimental vortex 4He I,first experimental vortex 4He II,Abrikosov vortex original,Experimentalsinglevortex1,experimentalvortex3}. In the current paper the starting point is a quantum system of spin-$0$ particles in the plane. This quantum system could in reality be, for example, a flat ordinary BEC or superfluid $^4\text{He}$-film. These systems can be described by an order parameter field taking complex values. The system under study has a global $U(1)$-symmetry and thus the first homotopy group, classifying pointlike topological defects is isomorphic to the additive group of integers. The vortices are pointlike objects characterized by a single integer, called the winding number or topological charge.

The pointlike nature of the vortices described above can be described as point particles in a well known way (for example \cite{Hodgedecompositionandthevortexhamiltonian,Observationsonthedynamicsofthetwo-dimensionalvortexgasoncompactRiemannsurfaces}). We will use this particle-description. As a consequence the effects of the vortex core are left out of the discussion. This is an approximation that can be justified when the vortex size of the vortex core is small compared to the dimensions of the whole physical system and the distances between vortices. This is in many actual cases true, and thus describing planar vortices as point particles presents a rather good approximation of real planar vortices. 

Our main interest is in a homogeneous system confined in an annular region (ring). We will incorporate the appropriate boundary conditions by using a simple Green's function method arising from the hydrodynamical description of the gas (eg. \cite{Kirchoff-routhinmultiplyconnecteddomains,Themotionofvorticesaroundmultiplecircularislands}). The choice of an annular region as the object of study is motivated by two reasons. First, the annulus is the most simple doubly connected domain in $\reals^2$ and the Green's function needed in our analysis is quite easily expressed analytically \cite{courantandhilbert}. This enables simple numerics and also some analytic observations of vortex behaviour inside the annulus. Also, using the theory of conformal mappings, the annulus can be mapped to other doubly connected regions in $\reals^2$. The second reason, more practical in nature, is that a ring shaped BEC can be produced by creating the condensate in a trap characterised by the so called ''mexican hat'' -potential. This means that experimental studies of vortex behaviour in an annular BEC are possible.

The dynamics of point vortices in an annulus has previously been studied in \cite{motionofvorticesintheannulus}. We obtain the same results along with new results. We will show that there always exists stationary (though not energetically stable) one- and two-vortex solutions in the annulus. These solutions are characterized by specific values of $r$ (the distance from the origin in polar coordinates) which we will call the stationarity radii. These radii divide the annulus into two subsets in the radial direction in such a way that position of the vortices with respect to the stationarity radii greatly affects the characteristics of a particular vortex solution. This fact has, at least to our knowledge, been unnoted before and is responsible for the new solutions that are not mentioned in \cite{motionofvorticesintheannulus}. The stationarity radii can, in general, be seen as relatives of the equilibrium points (or critical points) of the Green's function in the annulus, studied in eg. \cite{The equilibrium point for the green's function in the annulus,Equilibrium point of greens function and the eisenstein series}. They are however not the same thing, because the calculation of the stationarity radii takes into account the self-energies of the vortices. These self-energies are associated with the harmonic part of the two-dimensional Green's function.

Demonstrating the importance of the stationarity radii, we find analytic solutions describing corotating (the angular velocities of vortices have the same sign) and counterrotating (the angular velocities of vortices have opposite signs) motion for both vortex-vortex and vortex-antivortex cases. These solutions could not exist without the existence of the stationarity radii. With numerical simulations we further demonstrate how the stationarity radii along with the initial positions of the vortices play an essential role in the characteristics of the solutions of vortex motion. 

\section{Vortices as elementary particles in a planar quantum system}

\subsection{Description of planar vortices as point particles}

We will briefly review a simple way by which the point-particle theory of planar vortices (e.g. \cite{Hodgedecompositionandthevortexhamiltonian,Observationsonthedynamicsofthetwo-dimensionalvortexgasoncompactRiemannsurfaces,vortexconservationlaws}) can be derived by separating the vortices from the non-vorticical fluid part. The starting point is the semiclassical Lagrangian for a two-dimensional superfluid consisting of bosonic, spinless particles 
\begin{align}
L[\phi,\phi^*] = \int \diff^2 \! \vec x \Big( & \frac{i}{2}(\phi^* \partial_0 \phi-\partial_0 \phi^* \phi) - \frac{1}{2} |\nabla \phi|^2 \nonumber \\ & - U(\vec x) \phi^* \phi - V(\phi^* \phi) \Big).
\end{align}
The function $U$ represents external potentials and $V$ describes particle-particle interactions. Instead of the fields $\phi$ and $\phi^*$, the gas can also be described by hydrodynamical variables $\rho$ and $\vec v$ which are the particle density of the fluid and the velocity field of the particles. 

We will next choose the following parametrization, a two-dimensional analogue of the usual Helmholtz-decomposition of three-dimensional vector fields, for the velocity field $\vec v$:
\begin{equation}\label{iobfcwoiub}
v^i = \partial^i \varphi + \varepsilon^{ij} \partial_j A.
\end{equation}
Function $\varphi$ corresponds to the regular part of the phase of $\phi$ while the function $A$ is associated with the irregular (singular) part of the phase. The familiar quantization condition for $N$ pointlike vortices with winding numbers $\{q^1, \ldots q^N\}$ in the system leads to a simple expression 
\begin{equation}\label{ibeobcoqubcy}
A(\vec x) = - \sum_{a=1}^N q^a G(\vec x, \vec x^a),
\end{equation}
where the two-dimensional Green's function $G$ has the form
\begin{equation}
G(\vec x,\vec x^a) = \frac{1}{2 \pi} \log |\vec x - \vec x^a| + \tilde{G}(\vec x,\vec x^a).
\end{equation}
The harmonic function $\tilde{G}$ is determined by the requirement that there is no flow in or out of the system. This implies the Dirichlet type boundary condition $G=\text{const.}$ along the boundary or boundaries of the system.

After inserting $\varphi$ and $A$ into the hydrodynamical Lagrangian and forming the equations of motion from the usual action principle, we obtain the following Lagrangian describing the point vortex system
\begin{equation}
L^{\text{v}} = L^{\text{vortex}} + L^{\text{int}} + L^{\text{pot}},
\end{equation}
where the vortex system $L^{\text{vortex}}$, interaction $L^{\text{int}}$ and irrotational $L^{\text{pot}}$ parts are
\begin{eqnarray}
L^{\text{vortex}} & = & \frac{1}{2} \sum_a q^a \rho(\vec x^a) \varepsilon^{ij} \dot{x^a}_i x^a_j  \nonumber \\ & & - \frac{1}{2} \sum_a (q^a)^2 \rho(\vec x^a) \tilde{G}(\vec x^a) \nonumber \\ & & - \frac{1}{2} \sum_{a \neq b} q^a q^b \rho(\vec x^a) G(\vec x^a,\vec x^b), \\
L^{\text{int}} & = & \sum_a \int \diff^2 \! \vec x \, q^a J^{\text{pot}}_i \varepsilon^{ij} \partial_j G(\vec x,\vec x^a) \\
L^{\text{pot}} & = & L[\rho,\varphi] \label{bbbbbb},
\end{eqnarray}
where $\tilde{G}(\vec x^a) \equiv \tilde{G}(\vec x^a,\vec x^a)$ and $\vec{J}^{\text{pot}}$ is the part of the current density associated with irrotational flows. The relation of the Green's function to the vortex dynamics obtained here is similar to the results of \cite{Kirchoff-routhinmultiplyconnecteddomains}.

The vortices interact with each other through a potential function given by the appropriate Green's function $G$. In (\ref{bbbbbb}) we have obtained the self-energy of a single vortex, which is the energy associated with the flows created in the system by the presence of the vortex, by means of a simple renormalization procedure. The renormalization is performed by removing the singular logarithmic part of the Green's function from $L^{\text{vortex}}$ when the indices $a$ and $b$ coincide. In this coreless treatment the usual ultraviolet divergence present when calculating the vortex energy by direct integration is taken care of by this renormalization procedure. We see that the kinetic energy (or self energy) of a vortex is proportional to $\tilde{G}(\vec x)$ evaluated at the position of the vortex. This is a direct concequence of the defining properties of the Green's function. 

The equations of motion for the vortices are simply
\begin{equation}\label{ihyubfoe}
\rho(\vec x^a) \dot{\vec x}^a = \vec{J}(\vec x^a), \quad a=1,\ldots N.
\end{equation}
The study of vortex motion thus reduces to the hydrodynamical problem of studying the flows in the system. When non-vorticical flows are absent, the problem is to find the suitable Green's function under the Dirichlet boundary conditions defined above and then solve a group of nonlinear differential equations defined by (\ref{ihyubfoe}) above. 

\section{Vortex dynamics in the Annulus}

\subsection{Vortex Hamiltonians in the annulus}

From this point on in the current paper, we will assume that the particle density $\rho$ takes a constant value $\rho_0$ everywhere. It should be noted that this fixing of $\rho$ along with the requirement of energy conservation means that there is no energy transfer between the vortex system and the rest of the fluid. This also means that vortex-antivortex annihilation and other processes that affect the vortex configuration are not possible in our model. The vortices initially placed in the system will exist forever.

The form for the Green's function in the annulus can be found eg. in \cite{courantandhilbert}. It can be expressed as
\begin{align}\label{ibufeoiefbc}
G(r_1,\theta_1,r_2,\theta_2) = & - \frac{1}{2 \pi} \frac{\log r_1 \log r_2}{\log \eta} \nonumber  \\ & \!\!\!\!\!\!\!\!\!\!\!\!\!\!\!\!\!\!\!\!\!\!\!\!\!\!\!\!\!\!\!\!\!\!\!\!\!\!\!\!\!\!\!\!\!\!\! + \frac{1}{2 \pi} \log \left| \frac{\vartheta_1(i\log(r_1/r_2)/2 - (\theta_1-\theta_2)/2,\eta)}{\vartheta_4(i \log(r_1 r_2)/2 + (\theta_1-\theta_2)/2,\eta)} \right|,
\end{align}
where the $\vartheta_i$ are the Jacobi theta fuctions. Here we have chosen the inner and outer radius of the annulus to be $R_1=\eta^{1/2}$ and $R_2=\eta^{-1/2}$, respectively, so that the ratio of the inner radius to the outer radius is given by the parameter $0<\eta<1$ appearing in the theta functions. There is no fear of confusion since the $R_i$s can be removed and reinserted easily using their relation to $\eta$ and dimensional analysis. An important quantity is the geometric mean of the inner and outer circle bounding the annulus. It has the value 
\begin{equation}
\sqrt{R_1 R_2} = 1,
\end{equation}
when expressed in terms of $\eta$ defined above.

Since we are interested in the dynamics of one- and two-vortex systems, we write the two-vortex Hamiltonian $H^2$ in a form that is easily derived from the above Green's function
\begin{widetext}
\begin{align}
H^2 =\frac{q_1 q_2 \rho_0}{2 \pi} \bigg[ & - \log | \vec r_1 - \vec r_2 | - \frac{\log(r_1 \eta^{1/2}) \log(r_2 \eta^{1/2})}{\log \eta} + \frac{\log \eta}{2} \nonumber \\ & + 2 \sum_{n=1}^{\infty} \frac{\eta^n \cos(n(\theta_1-\theta_2))}{n(1-\eta^{2n})} \bigg( (r_1 r_2)^n+\frac{1}{(r_1 r_2)^n} - \eta^n \bigg( \frac{r_2^n}{r_1^n}+\frac{r_1^n}{r_2^n} \bigg) \bigg) \bigg] + \sum_{a=1,2} H^1_a,
\end{align}
\end{widetext}
where the $H^1_a$ are the radially symmetric renormalized single-vortex Hamiltonians given by
\begin{widetext}
\begin{align}\label{svHamiltonian}
H^1_a = \frac{q_a^2 \rho_0}{4 \pi} \bigg[ -\frac{\log^2 (r \eta^{1/2})}{\log \eta} + \frac{\log \eta}{2} + 2 \sum_{n=1}^{\infty} \frac{\eta^n}{n(1-\eta^{2n})} \Big( r^{2n}+\frac{1}{r^{2n}} \Big) - 4 \sum_{n=1}^{\infty} \frac{\eta^{2n}}{n(1-\eta^{2n})} \bigg].
\end{align}
\end{widetext}
We will use the above Hamiltonians in this and the next section to study one- and two-vortex systems in the annulus.

\subsection{Single vortex in an annulus}

\begin{figure} 
	\includegraphics[angle=0,width=0.35\textwidth]{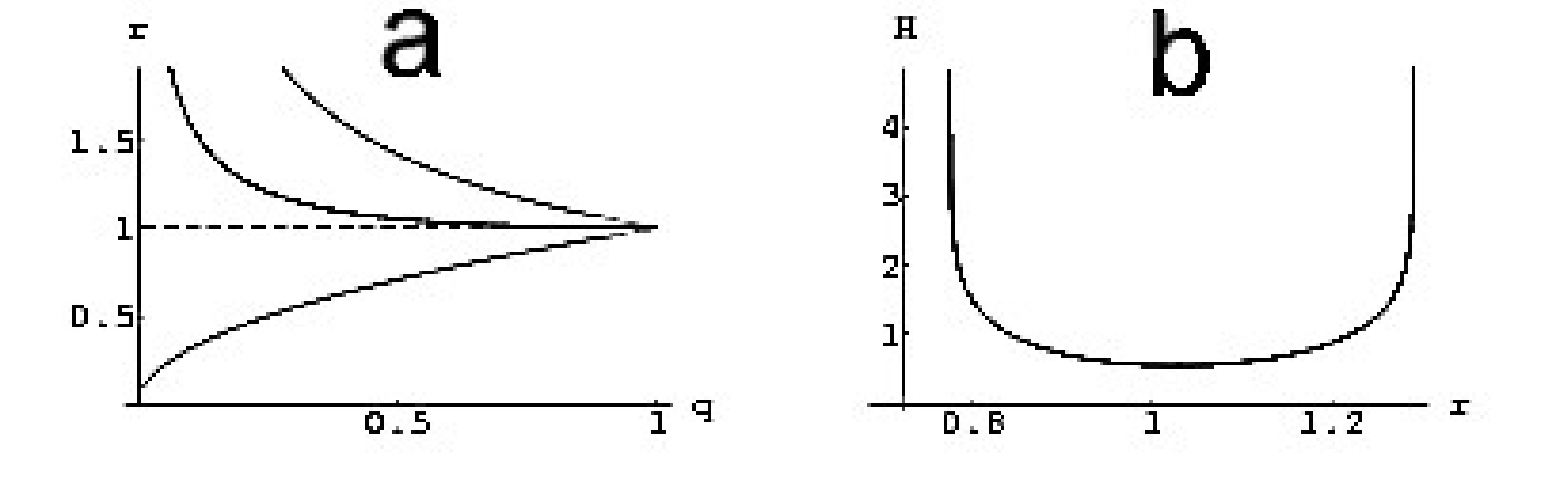}
\caption{Figure (a) shows $R_1$, $R_2$ and between them the stationarity radius $R_{\text{S}}$ as a function of $q=R_1/R_2$, all approaching the dashed line $r=1$ as $q \rightarrow 1$. Figure (b) shows a typical graph of the single-vortex energy as a function of the distance from the center of the annulus.} \label{Vorteksikuva32.eps}
\end{figure}

The behaviour of a single vortex is obtained simply by studying the single-vortex Hamiltonian (\ref{svHamiltonian}). The result is a vortex moving with the renormalized flows it has created in the annulus
\begin{align}
\vec{v}^{\text{self}} = & \frac{q}{2 \pi} \Big[ \frac{\log (r \eta^{1/2})}{r \log \eta} \nonumber \\ & - \frac{1}{r} \sum_{n=1}^\infty \frac{1}{1-\eta^{2n}} \Big( \frac{ (r \eta^{1/2})^{4n} - \eta^{2n}}{(r \eta^{1/2})^{2n}} \Big) \Big] \hat{\vec e}_{\theta}.
\end{align}
This implies the well known fact that a single vortex in an annulus will move along a circle at a constant distance from the origin \cite{motionofvorticesintheannulus}. We see that there is always a  single value of $r$ such that the vortex will remain at rest on the circle defined by that value. We will call this value of $r$ the (single-vortex) stationarity radius and denote it with $R_{\text{S}}$. It is easily seen that $R_{\text{S}} >1$ for all values of $\eta$. The approximate value of $R_{\text{S}}$ can easily be solved numerically (\ref{Vorteksikuva32.eps} (b)) for different $\eta$. From the form of the single-vortex Hamiltonian (\ref{svHamiltonian}) we see that the circle $r=R_{\text{S}}$ corresponds to a degenerate minimum value of the vortex energy (\ref{Vorteksikuva32.eps} (b)).

A single vortex cannot exist in the annulus if there is no angular momentum $L_z$ in the system. We must thus also take into account the energy associated with this angular momentum. It can easily be shown that the vortex configuration that minimizes the energy of the fluid, is the one that minimizes the free energy defined by
\begin{equation}
F = E - \omega L_z,
\end{equation}
where $\omega$ is the angular velocity, with which the annulus is being rotated \cite{Landau and Lifshitz Statistical Mechanics}. Since we are dealing with the system  in a classical manner, the moment of inertia has the (classical) rigid value
\begin{equation}
I = \frac{\partial L_z}{\partial \omega} = N \braket{r^2} = \frac{\pi \rho_0}{2} \frac{1-\eta^4}{\eta^2},
\end{equation}
where $N$ is the total number of particles in the gas \cite{Moment of Inertia and Superfluidity of a Trapped Bose Gas}. The angular momentum of a vortex at distance $r$ from the origin is easily calculated to be
\begin{equation}
L_z = \frac{q\rho_0}{2} \frac{1+\eta^2-r^2 \eta^2}{\eta} + \frac{q\rho_0}{4} \frac{1-\eta^2}{\eta \log\eta}.
\end{equation}
We will not get into the details of the equation for the minimum of the free energy. It suffices to note that, since $\omega L_z$ is a decreasing function of $r$, the minimum of free energy is located between $R_{\text{S}}$ and the outer radius of the annulus.

\section{vortex pairs in the annulus}

We next proceed to study the dynamics of two-vortex systems in the annulus analytically and numerically. In subsections A and B we study analytically vortex-vortex and vortex-antivortex pairs respectively. In subsection C we study the motion of vortex pairs in the annulus numerically and demonstrate the results of subsections A and B.

\subsection{A vortex-vortex pair in an annulus}

It is easy to see that when the condition $\theta_1-\theta_2=\pi$ is met, the motion of the vortex-vortex pair has simple analytic solutions for which $\dot{\theta}_1-\dot{\theta}_2=0$. These type of solutions are obtained choosing the initial positions for the vortices as $\theta_1(0)-\theta_2(0)=\pi$ and $r_1(0)=r_2(0)$. This corresponds to a vortex-vortex pair moving on the same circle with constant angular velocity. The angular velocity of the corotating pair is
\begin{align}\label{iwfbeirqbfeipr}
\dot{\theta}^{\text{corot}}_{q_1=q_2} & = \frac{q_1}{\pi} \bigg[ \frac{1}{r^2} \frac{\log(r)}{\log(\eta)} \nonumber \\ & + \frac{1}{r^2} \sum_{n=1}^\infty \frac{\eta^{2n}}{1-\eta^{4n}} \bigg( r^{4n}-\frac{1}{r^{4n}} \bigg) + \frac{1}{4} \frac{1}{r^2} \bigg]. 
\end{align} 
Setting $\dot{\theta}^{\text{corot}}_{q_1=q_2} = 0$ we see that there always exists a single solution for which the vortex pair remains at rest. This is analogous to the case of the single vortex. However, this time the stationary solution appears for a value $r<1$ for the radii of the vortices. We will call this radius the vortex-vortex stationarity radius and denote it with $R^{++}_{\text{S}}$ (or $R^{--}_{\text{S}}$, depending on the sign of the winding numbers). We have thus shown that for a vortex-vortex pair in an annulus there always exists infinitely many corotating solutions and also a stationary solution. The stationary solution minimizes the energy of the vortex pair. 

A solution satisfying the different condition $\dot{\theta}_1+\dot{\theta}_2=0$ is called counterrotating as the angular velocities of the vortices have opposite sign in this case. Such solution can appear when the vortices are on different sides of the stationarity radius; the stationarity radius corresponds to a solution which is both corotating and counterrotating. The counterrotating solutions are not as easily described analytically as the corotating solutions, but they are easily studied numerically.

As in the case of a single vortex, a vortex-vortex pair cannot exist in the annulus without the precense of angular momentum. The free energy is easily calculated since $L_z$ is linear with respect to the velocity fields and thus additive. Again, it suffices to note that the most favourable solutions are corotating solutions where both vortices are displaced (at least approximately) by the same amount from the stationarity circle towards the outer radius given the symmetry of the system and the complex nature of the vortex trajectories in solutions where $r_1$ and $r_2$ differ by a ''larger'' amount (we will not study the concept of ''large'' in this context here). This is in accordance with the results of \cite{Vortex signatures in annular BECs} for the stable vortex configurations in an annular BEC.

\subsection{A vortex-antivortex pair in an annulus}

Because of the change of sign in the interaction part of the Hamiltonian, the conditions $\theta_1-\theta_2=\pi$ and $r_1=r_2$ describe a counterrotating solution. Conversely, the solutions describing corotating behaviour are not easily expressed analytically but are easily studied numerically. Searching for a stationary solution leads us to study the equation
\begin{equation}\label{lkrahglku}
1 - 4 \sum_{n=1}^\infty \frac{\eta^{2 n - 1}}{1-\eta^{4 n - 2}} \frac{1-r^{8 n - 4}}{r^{4 n - 2}} = 0.
\end{equation}
This equation always has a solution $r<1$ and thus there always exists a stationary vortex-antivortex -solution, this solution being again the solution which is both corotating and counterrotating and correspomding to a local minimum of the vortex-antivortex energy. We will call this value of $r$ the vortex-antivortex stationarity radius and denote it with $R^{+-}_{\text{S}}$.

It may be noted that for the case of the vortex-antivortex pair, there need not be any angular momentum in the system. In this case the free energy is just the energy of the vortex pair. One way to produce a vortex-antivortex pair confined in an annular region is to produce a soliton of topological charge $0$ in a BEC confined in a ''mexican hat''-potential. This soliton can then decay, typically producing a vortex-antivortex pair on opposite sides of the annulus. Of course, such a soliton will carry with it some angular momentum. The situation still remains different from the case of one or two vortices of winding numbers of the same sign. The qualitative properties of the corotating and counterrotating solutions suggest that they both can be at least metastable energetically. A confirmation for this fact can be obtained through direct numerical simulations of a ring-shaped BEC starting from the Gross-Pitayevskii equation.

\subsection{Numerical simulations}

The equations of motion describing vortex pairs in the annulus can easily be solved numerically using a standard adaptive stepsize Runge-Kutta method. We will choose the value of $\eta$ to be $1/3$, since then the annulus is similar to that used in \cite{motionofvorticesintheannulus} (scaled by a factor of $\approx 1.1547$), where the inner radius was chosen to be $0.5$cm and the outer radius $1.5$cm. We notice that very important quantities, as expected, are the two-vortex stationarity radii $R^{++}_{\text{S}} < 1$ and $R^{+-}_{\text{S}} < 1$, which in our case have the numerical values $R^{++}_{\text{S}} \approx 0.74911$ and $R^{+-}_{\text{S}} \approx 0.888928$.

\begin{figure} 
	\includegraphics[angle=0,width=0.35\textwidth]{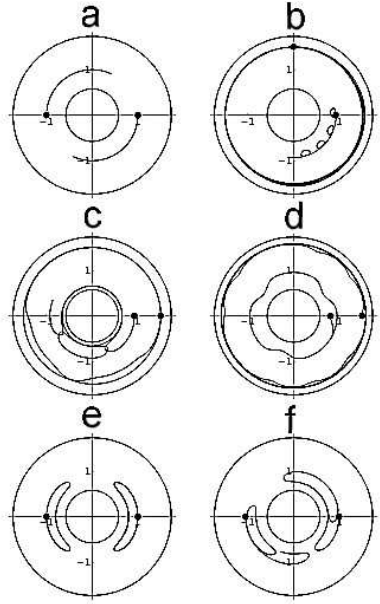}
\caption{(a) A corotating vortex-vortex pair. (b) Two vortices travelling in the clockwise direction with the outer one overtaking the inner one 4 times. (c) A vortex-antivortex pair moving in the same direction when the inner vortex is inside the circle $r=R^{+-}_{\text{S}}$ and in opposite directions when the inner vortex is outside the circle $r=R^{+-}_{\text{S}}$. (d) A vortex-antivortex pair moves in opposite directions, the outer one with a greater velocity. (e) A vortex and an antivortex with initial positions on the same circle perform motion along their own loops. (f) Is the the case of (e) with the initial position of the other vortex changed by a small amount in the radial direction, now the vortices have a net motion in the counterclockwise direction. Changing the initial position in the angle variable does not result in any net flows but produces a rotated version of figure (e). The initial positions of the vortices are denoted with the points in this and the following figures.}\label{Vorteksikuva42.eps}
\end{figure}

\begin{figure} 
	\includegraphics[angle=0,width=0.35\textwidth]{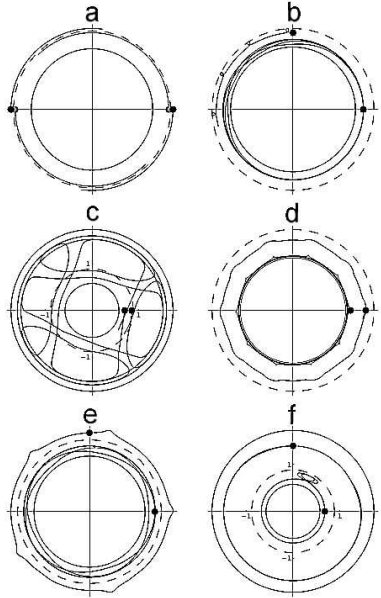}
\caption{(a) A vortex pair inside the dashed circle corotating in the counterclockwise direction (initial positions grey points) and outside the dashed circle corotating in the clockwise direction. (b) A vortex-vortex pair is inside the circle $r=R^{++}_{\text{S}}$. The inner vortex moves with greater angular velocity, otherwise the situation is analogous to figure \ref{Vorteksikuva42.eps} (b). (c) A vortex-antivortex pair starting inside the circle $r=R^{+-}_{\text{S}}$. Because of the closeness of their initial positions they both perform motion similar to the motion of the inner vortex of figure \ref{Vorteksikuva42.eps} (c) with the roles of the inner and outer boundary changed. (d) Another vortex-antivortex case with both vortices initially inside the circle $r=R^{+-}_{\text{S}}$. (e) A vortex-vortex pair with initial positions inside and outside the circle $r=R^{++}_{\text{S}}$ performs motion similar to the vortex-antivortex pair in figure \ref{Vorteksikuva42.eps} (d). (f) A vortex-antivortex pair with initial positions inside and outside the circle $r=R^{+-}_{\text{S}}$ performs motion similar to the vortex-vortex pair in figure \ref{Vorteksikuva42.eps} (b).}\label{Vorteksikuva52.eps}
\end{figure}

\begin{figure} 
	\includegraphics[angle=0,width=0.35\textwidth]{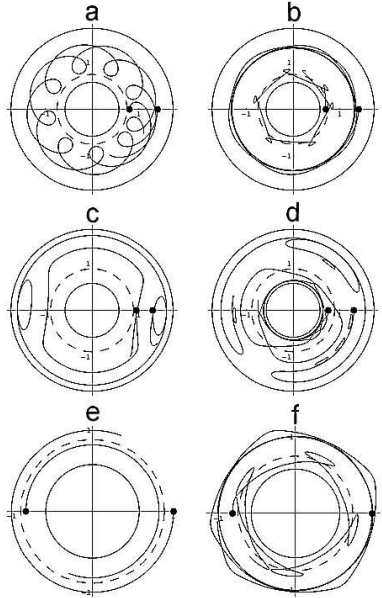}
\caption{In figures (a) and (b) we see the change in the behaviour of a vortex-vortex pair when the initial position of the inner vortex is moved from outside the circle $r=R^{++}_{\text{S}}$ to inside it. In (c) and (d) we see the corresponding change for a vortex-antivortex pair and the circle $r=R^{+-}_{\text{S}}$. (e) A corotating vortex-antivortex pair. (f) A vortex-vortex pair with the same initial positions as the vortex-antivortex pair in (e).}\label{Vorteksikuva62.eps}
\end{figure}

Figures \ref{Vorteksikuva42.eps} (a-e) are the same cases as in \cite{motionofvorticesintheannulus} in the figures 1-5, respectively. Figures \ref{Vorteksikuva42.eps} (e-f) show how sensitive the solutions are to initial conditions. In figures \ref{Vorteksikuva52.eps}  and \ref{Vorteksikuva62.eps} we see the effect of the stationarity radii for the two-vortex solutions. The dashed circle in these figures is the circle $r=R^{++}_{\text{S}}$ for a vortex-vortex pair and the circle $r=R^{+-}_{\text{S}}$ for a vortex-antivortex pair.

\subsection{A brief heuristic discussion}

The fact that we have been able to find analytic and numerical solutions with similar properties for both kinds of vortex pairs is easily explained by the geometry of the annulus. The annulus is not simply connected and the annulus has two boundaries. The first property has a profound consequence most simply illustrated by the vortex-vortex pair. We may think that the interaction between vortices is transmitted by the flows of the original system (BEC, superfluid or something else), or more simply that the interaction is transmitted only where the system exists ($\rho \neq 0$), in this case the annulus. Thus we may think that vortex $1$ interacts with vortex $2$ through two different routes, ''striking'' at vortex $2$ from both directions $\theta \rightarrow \theta_2-$ and $\theta \rightarrow \theta_2+$. With this thought in mind it is not surprising that vortices on exactly opposite sides of the origin corotate or stay still, the interactions from $\theta-$ and $\theta+$ essentially cancel each other and all that is left is a self-induced velocity for the vortex pair.

The two boundaries of the annulus have the following consequence: The harmonic part of the Green's function must satisfy the symbolic relation $\tilde{G}(\vec x^1,\vec x^1)= \tilde{G}(\vec{x}^1) = \infty$ on both boundaries. This and the finiteness of $\tilde{G}$ inside the annulus directly implies that $\tilde{G}(\vec{x}^1)$ must have a minimum value that corresponds to a stationary vortex solution. Because of the radial symmetry of the annulus this solution is infinitely degenerate and, as seen before, corresponds to a single circle cosentric to the circles bounding the annulus. The existence of this circle leads to the fact that the sign of the angular velocity of a vortex around the origin depends on whether it is inside or outside the stationarity circle. Adding more vortices to the system does not change the fact that the single vortex energies become infinite at the boundaries and that this leads to the existence of a stationarity radii. Thus vortex-vortex -pairs and vortex-antivortex -pairs can display similar behaviour.

\section{Conclusions}

We have shown that taking into account the boundary conditions of the doubly connected annulus leads to the existence of stationarity radii which greatly affect vortex motion in the annulus. This leads to the existence of corotating and counterrotating as well as stationary solutions for both vortex-vortex and vortex-antivortex pairs. 

\section*{Acknowledgments}

The author acknowledges the financial support of the Academy of Finland (grants no. 206108 and 115682), the Finnish Society of Sciences and Letters, Magnus Ehrnrooth Foundation and the Finnish Cultural Foundation.

\end{document}